\documentclass{article}
\usepackage{graphicx}
\RequirePackage{feynarts}
\newcommand{\epe}{\epsilon^\prime/\epsilon}
\newcommand{\be}{\begin{equation}}
\newcommand{\ee}{\end{equation}}
\newcommand{\bea}{\begin{eqnarray}}
\newcommand{\eea}{\end{eqnarray}}
\newcommand{\bra}{\langle}
\newcommand{\ket}{\rangle}
\newcommand{\pkppi}{p_K \cdot p_\pi}

\newcommand{\etal}{{\it et al}.\ }
\begin{document}
\rightline{BNL-HET-02/8}

\begin{center}

{\large\bf Lattice extraction of $ K \rightarrow \pi \pi $
amplitudes to $ O(p^{4}) $ in chiral perturbation theory }

\vspace{.2in}

Jack Laiho$^{*}$\\
 \noindent Department of Physics, Princeton
University, Princeton, NJ\ \ 08544\\
\smallskip
 Amarjit Soni$^{\dag}$ \\
\noindent Theory Group, Brookhaven National Laboratory, Upton, NY\
\
11973\\
\footnotetext{$^\dag$email: jlaiho@viper.princeton.edu\hskip1.5in
$^*$email: soni@bnl.gov}
\end{center}


\begin{quote}
We show that the lattice calculation of $K\to\pi\pi$ and $\epe$
amplitudes for (8,1) and (27,1) operators to $O(p^4)$ in chiral
perturbation theory is feasible when one uses $K\to\pi\pi$
computations at the two unphysical kinematics allowed by the
Maiani-Testa theorem, along with the usual (computable) two- and
three-point functions, namely $K\to0$, $K\to\pi$ (with momentum)
and $K$-$\bar K$. Explicit expressions for the finite logarithms
emerging from our $O(p^4)$ analysis to the above amplitudes are
also given.
\end{quote}
\section{Introduction and Motivation}

    Recent lattice QCD calculations done by the CP-PACS \cite{noaki}
    and RBC \cite{blum}  collaborations
    using domain wall fermions have made significant progress in explaining
the $ \Delta
    I = 1/2 $ rule in the decay $ K \rightarrow \pi \pi $, though
    their results for the direct \textit{CP} violation
    parameterized by
    $ \textrm{Re}(\epsilon'/\epsilon) $ differs rather drastically from experiment.
    Recall that  measurements at CERN \cite{fanti} and Fermilab \cite{alavi}
    have yielded an experimental grand average of
    $ \textrm{Re}(\epsilon'/\epsilon) = (17.2\pm1.8)\times 10^{-4} $.
   On the theoretical side, both  lattice collaborations
   find  a value of $ \textrm{Re}(\epsilon'/\epsilon)  \simeq -5\times
10^{-4}
   $, a \emph{negative} value, though both groups have made rather
   severe (uncontrolled) approximations.  Given that large cancellations
   occur between contributions of the strong and the electroweak
   penguins towards $\epe$ (cancellations that are not relevant in the calculation
   of the \textit{CP}-conserving $K\to\pi\pi$ amplitudes), and given the serious
   approximations, the disagreement with
   experiment for $\epe$ should not be totally unexpected\cite{pk,gm1}.

    One of these uncontrolled approximations was the use of the
   quenched approximation, where the fermion determinant in the
   path integral is set to 1 in order to make the problem
   tractable on current computers.  Another was the use of leading
   order chiral perturbation theory to relate unphysical $
   K \rightarrow \pi $ and $ K \rightarrow | 0 \rangle $ amplitudes to
   the physical $ K \rightarrow \pi \pi $ amplitudes.  This method
   was first proposed by Bernard, \etal \cite{bern}.  Because of the
   difficulty of extracting multi-hadron decay amplitudes from the
   lattice, as expressed by the Maiani-Testa theorem \cite{maiani}, it is
much
   easier to compute the two- and three-point functions
  (i.e. $K \rightarrow | 0 \rangle $ and $K \rightarrow \pi$,
  respectively) and use chiral
   perturbation theory (ChPT) to extrapolate to the physical
   matrix elements.

   It is likely, however, that next-to-leading order ChPT will
   introduce significant corrections ($ \sim 30\% $ or more) to the
   leading order amplitudes. Furthermore, since final state (strong)
   phases cannot
   arise at tree-level in the chiral amplitudes, chiral-loop
   corrections are essential to enable us to use the measured phases
   for the $I=0$ and 2 final states as additional testing ground of the
   calculational apparatus.  Unfortunately, at higher orders in ChPT the
   number of free parameters that enter the theory (and must be
   determined from first principle methods like the lattice)
   proliferates rapidly.  It has been shown by Cirigliano and
   Golowich \cite{cirig} that the dominant electroweak penguin contributions
  [ $ (8_{L},8_{R})'s $ ] to $K\to\pi\pi$ can be recovered at next-to-leading
order (NLO)
  from $ K \rightarrow \pi $ amplitudes using momentum insertion.
   Bijnens \etal \cite{bijnens} showed how to obtain most of the low
  energy constants (LEC's) relevant for the case of the
 $ (8_{L},1_{R})'s  $ and $ (27_{L},1_{R})'s  $ using off-shell $ K
\rightarrow \pi
  $  Green's functions;  not all LEC's could be determined,
  however.

On the lattice, though, not only $K\to|0\rangle$ and $K\to\pi$
with momentum insertion are calculable, but so is $K\to\pi\pi$ at
the two values of unphysical kinematics for which the Maiani-Testa
theorem can be bypassed. To recapitulate, despite Maiani-Testa
restrictions, direct calculation of $K\to \pi\pi$ on the lattice
is accessible at (i) $m^{\rm lattice}_K = m^{\rm lattice}_\pi$
where the weak operator inserts energy \cite{berntwo} and (ii)
$m^{\rm lattice}_K=2m^{\rm lattice}_\pi$, i.e.\ at threshold
\cite{dawson}. We will refer to these two special locations as
unphysical kinematics point 1 (UK1) and point 2 (UK2),
respectively. In this work we therefore focus on using $ K
\rightarrow | 0 \rangle $, $ K \rightarrow \pi
  $ with momentum insertion and $K$-$\bar K$, along with information from
  $ K \rightarrow \pi \pi $ at these
  two unphysical values of the kinematics which are accessible to  the
lattice\cite{spqcdr}.
  Thereby, we are able to  show that all the
  relevant $ O(p^{4}) $ LEC's can be recovered for $ K \rightarrow \pi \pi
  $ in the physical $ (8_{L},1_{R})$ and $ (27_{L},1_{R})$ cases.
Expressions for $O(p^4)$ finite logarithimic contributions to all
the processes that may be needed for fitting the lattice data are
then given.

  The content of the paper is as follows.  Section 2 very briefly
  recapitulates the
  formalism of effective four-fermion operators in a Standard Model
  calculation.  Section 3 reviews ChPT and the realization of the
  effective four-quark operators in terms of ChPT operators for
  weak processes.  Section 4 presents the results of this paper,
  showing how to obtain the low energy constants necessary for
  physical $K\to\pi\pi$ amplitudes at $O(p^4)$ from quantities
  which can, in principle, be computed on the lattice.  Section 5
  presents the conclusion.  Finally, the finite logarithmic contributions to
  the relevant amplitudes are presented in a set of Appendixes.

\section{Effective Four Quark Operators}

In the Standard Model, the nonleptonic interactions can be
expressed in terms of an effective $ \Delta S=1 $ hamiltonian
using the operator product expansion \cite{ciuch,bucha},

\begin{equation}\label{1}
    \langle \pi \pi |{\cal H}_{\Delta S=1}|K\rangle =
    \frac{G_{F}}{\sqrt{2}}V_{ud} V^{*}_{us} \sum c_{i}(\mu)
    \langle \pi \pi|Q_{i}|K\rangle_{\mu},
\end{equation}

\noindent where $ c_{i}(\mu) $ are the Wilson coefficients
containing the short distance perturbative physics, and the matrix
elements $ \langle \pi \pi|Q_{i}|K\rangle_{\mu} $ must be
calculated nonperturbatively.  The four quark operators are

\begin{equation}\label{2}
    Q_{1}=\overline{s}_{a} \gamma_{\mu} (1-\gamma^{5}) u_{a}
    \overline{u}_{b}\gamma^{\mu} (1-\gamma^{5}) d_{b},
\end{equation}
\begin{equation}
Q_{2}=\overline{s}_{a} \gamma_{\mu} (1-\gamma^{5}) u_{b}
    \overline{u}_{b}\gamma^{\mu} (1-\gamma^{5}) d_{a},
\end{equation}
\begin{equation}
Q_{3}=\overline{s}_{a} \gamma_{\mu} (1-\gamma^{5}) d_{a} \sum_{q}
    \overline{q}_{b}\gamma^{\mu} (1-\gamma^{5}) q_{b},
    \end{equation}
    \begin{equation}
Q_{4}=\overline{s}_{a} \gamma_{\mu} (1-\gamma^{5}) d_{b} \sum_{q}
    \overline{q}_{b}\gamma^{\mu} (1-\gamma^{5}) q_{a},
    \end{equation}
     \begin{equation}
Q_{5}=\overline{s}_{a} \gamma_{\mu} (1-\gamma^{5}) d_{a} \sum_{q}
    \overline{q}_{b}\gamma^{\mu} (1+\gamma^{5}) q_{b},
     \end{equation}
     \begin{equation}
Q_{6}=\overline{s}_{a} \gamma_{\mu} (1-\gamma^{5}) d_{b} \sum_{q}
    \overline{q}_{b}\gamma^{\mu} (1+\gamma^{5}) q_{a},
    \end{equation}
    \begin{equation}
Q_{7}=\frac{3}{2} \overline{s}_{a} \gamma_{\mu} (1-\gamma^{5})
d_{a}
     \sum_{q} e_{q}
    \overline{q}_{b}\gamma^{\mu} (1+\gamma^{5}) q_{b},
    \end{equation}
    \begin{equation}
Q_{8}=\frac{3}{2} \overline{s}_{a} \gamma_{\mu} (1-\gamma^{5})
d_{b}
     \sum_{q} e_{q}
    \overline{q}_{b}\gamma^{\mu} (1+\gamma^{5}) q_{a},
     \end{equation}
      \begin{equation}
Q_{9}=\frac{3}{2} \overline{s}_{a} \gamma_{\mu} (1-\gamma^{5})
d_{a}
     \sum_{q} e_{q}
    \overline{q}_{b}\gamma^{\mu} (1-\gamma^{5}) q_{b},
    \end{equation}
    \begin{equation}
Q_{10}=\frac{3}{2} \overline{s}_{a} \gamma_{\mu} (1-\gamma^{5})
d_{b}
     \sum_{q} e_{q}
    \overline{q}_{b}\gamma^{\mu} (1-\gamma^{5}) q_{a}.
\end{equation}

In the effective theory $ Q_{1} $ and $ Q_{2} $ are the
current-current weak operators, $ Q_{3}-Q_{6} $ are the operators
arising from QCD penguin diagrams, while $ Q_{7}-Q_{10} $ are the
operators arising from electroweak penguin diagrams.

\section{Chiral Perturbation Theory}

    Chiral perturbation theory (ChPT) is an effective quantum
    field theory where the quark and gluon degrees of freedom have
    been integrated out, and is expressed only in terms of the
    lowest mass pseudoscalar mesons \cite{georgi}.  It is a perturbative
    expansion about small quark masses and small momentum of the low mass
    pseudoscalars.  The effective Lagrangian is made up of
    complicated nonlinear functions of the pseudoscalar fields,
    and is nonrenormalizable, making it necessary to introduce
    arbitrary constants at each order in perturbation theory.  In
    such an expansion, operators of higher order in the momentum
    (terms with increasing numbers of derivatives) or mass appear at
    higher order in the perturbative expansion.  The most general
    set of operators at a given order can be constructed out of
    the unitary chiral matrix field $ \Sigma $, given by

    \begin{equation}\label{3}
    \Sigma = \exp \left[\frac {2i\phi^{a}\lambda^{a}}{f}\right],
\end{equation}

    \noindent where $ \lambda^{a} $ are proportional to the Gell-Mann
    matrices with $tr(\lambda_a\lambda_b)=\delta_{ab}$, $ \phi^{a} $ are the
real pseudoscalar-meson fields,
    and $ f $ is the meson decay constant in the chiral limit, with $ f_{\pi} $
    equal to 130 MeV in our convention.

    At leading order $ [O(p^{2})] $ in ChPT, the strong Lagrangian
    is given by

    \begin{equation}\label{4}
    {\cal L}^{(2)}_{st}=\frac{f^{2}}{8}
    \textrm{tr}[\partial_{\mu}\Sigma\partial^{\mu}\Sigma] +
    \frac{f^{2}B_{0}}{4}\textrm{tr}[\chi^{\dag}\Sigma+\Sigma^{\dag}\chi],
 \end{equation}

\noindent where $ \chi= (m_u,m_d,m_s)_{\rm diag} $ and

 $ B_{0}= \frac{m^{2}_{\pi^{+}}}{m_{u}+m_{d}}=
\frac{m^{2}_{K^{+}}}{m_{u}+m_{s}}=\frac{m^{2}_{K^{0}}}{m_{d}+m_{s}}.\\
$

\noindent The leading order weak chiral Lagrangian is given by
\cite{golt}

\begin{eqnarray}\label{5}
    {\cal L}^{(2)}_{W}&  =  &\alpha_{88}\textrm{tr}[\lambda_{6}\Sigma Q \Sigma^{\dag}]
               +
    \alpha_{1}\textrm{tr}[\lambda_{6}\partial_{\mu}\Sigma\partial^{\mu}\Sigma^{\dag}]
    +\alpha_{2} 2B_{0}
    \textrm{tr}[\lambda_{6}(\chi^{\dag}\Sigma+\Sigma^{\dag}\chi)] \nonumber \\
    \!\!& & +
    \alpha_{27}t^{ij}_{kl}(\Sigma\partial_{\mu}\Sigma^{\dag})^{k}_{i}
    (\Sigma\partial^{\mu}\Sigma^{\dag})^{l}_{j} + \textrm{H.c.},
\end{eqnarray}

\noindent where $t^{ij}_{kl}$ is symmetric in $i, j$ and $k, l$,
traceless on any pair of upper and lower indices with nonzero
elements $t^{13}_{12}=1$, $t^{23}_{22}=1/2$ and
$t^{33}_{32}=-3/2$. Also, $ Q $ is the quark charge matrix, $
Q=1/3(2,-1,-1)_{\rm diag}$
 and $ (\lambda_6)_{ij}=
 \delta_{i3}\delta_{j2}$.

The terms in the weak Lagrangian can be classified according to
their chiral transformation properties under
$\textrm{SU}(3)_{L}\times \textrm{SU}(3)_{R}$.  The first term in
(14) transforms as $ 8_{L}\times 8_{R} $ under chiral rotations
and corresponds to the electroweak penguin operators $ Q_{7} $ and
$ Q_{8} $.  The next two terms in (14) transform as $ 8_{L}\times
1_{R} $, while the last transforms as $ 27_{L}\times 1_{R} $ under
chiral rotations.  All ten of the four quark operators of the
effective weak Lagrangian have a realization in the chiral
Lagrangian differing only in their transformation properties and
the values of the low energy constants which contain the
non-perturbative dynamics of the theory.

For the transition of interest, $ K \rightarrow \pi \pi $, the
operators can induce a change in isospin of $\frac12$ or $\frac32$
depending on the final isospin state of the pions.  We can then
classify the isospin components of the four quark operators
according to their
transformation properties \cite{noaki,blum}:\\

\hspace{1cm}\noindent $ Q^{1/2}_{1}, Q^{1/2}_{2}, Q^{1/2}_{9},
Q^{1/2}_{10}:
8_{L}\times 1_{R} \oplus 27_{L}\times 1_{R}$;\\

\hspace{1cm}\noindent $Q^{3/2}_{1}, Q^{3/2}_{2}, Q^{3/2}_{9},
Q^{3/2}_{10}:  27_{L}\times 1_{R}$;\\

\hspace{2cm}\noindent $Q^{1/2}_{3}, Q^{1/2}_{4}, Q^{1/2}_{5},
Q^{1/2}_{6}: 8_{L}\times
1_{R}$;\\

\hspace{2cm}\noindent$Q^{1/2}_{7}, Q^{1/2}_{8}, Q^{3/2}_{7},
Q^{3/2}_{8}: 8_{L}\times
8_{R}.\\\\
$

Note that $ Q_{3}-Q_{6} $ are pure isospin $\frac12$ operators.
This paper deals only with the $ 27_{L}\times 1_{R} $ and $
8_{L}\times 1_{R} $ operators.  For the treatment of the $
8_{L}\times 8_{R} $ operators to $ O(p^{2}) $ NLO see Ref.
\cite{cirig}.  At NLO the strong Lagrangian involves 12 additional
operators with undetermined coefficients.  These were introduced
by Gasser and Leutwyler in \cite{gass}.  The complete basis of
counterterm operators for the weak interactions with $ \Delta S=1,
2 $ was treated by Kambor, Missimer and Wyler in \cite{kambor} and
\cite{ecker}. A minimal set of counterterm operators contributing
to $ K \rightarrow \pi $ and $ K \rightarrow \pi \pi $ for the $
(8_{L},1_{R})$ and $ (27_{L},1_{R})$ cases is given by
\cite{golt}, with the effective Lagrangian

\begin{equation}\label{7}
    {\cal L}^{(4)}_{W}= \sum e_{i} {\cal O}^{(8,1)}_{i}+ \sum
    d_{i}{\cal O}^{(27,1)}_{i},
\end{equation}

\begin{equation}
\begin{array}{ll}
 {\cal O}^{(8,1)}_{1}= \textrm{tr}[\lambda_{6} S^2], & {\cal O}^{(27,1)}_{1}=t^{ij}_{kl}
(S)^{k}_{i}(S)^{l}_{j},\\
{\cal O}^{(8,1)}_{2}= \textrm{tr}[\lambda_{6} S] \textrm{tr}[S], &
{\cal O}^{(27,1)}_{2}=t^{ij}_{kl}
(P)^{k}_{i}(P)^{l}_{j},\\
{\cal O}^{(8,1)}_{3}=\textrm{tr}[\lambda_{6} P^{2}], &
{\cal O}^{(27,1)}_{4}=t^{ij}_{kl}(L_{\mu})^{k}_{i}(\{L^{\mu},S\})^{l}_{j}, \\
{\cal O}^{(8,1)}_{4}=\textrm{tr}[\lambda_{6} P] \textrm{tr}[P], &
{\cal O}^{(27,1)}_{5}=t^{ij}_{kl}
(L_{\mu})^{k}_{i}([L^{\mu},P])^{l}_{j},\\
{\cal O}^{(8,1)}_{5}=\textrm{tr}[\lambda_{6}[S,P]], & {\cal
O}^{(27,1)}_{6}=t^{ij}_{kl}
(S)^{k}_{i}(L^{2})^{l}_{j},\\
{\cal O}^{(8,1)}_{10}=\textrm{tr}[\lambda_{6} \{S,L^{2}\}], &
{\cal O}^{(27,1)}_{7}=t^{ij}_{kl}
(L_{\mu})^{k}_{i}(L^{\mu})^{l}_{j}
\textrm{tr}[S],\\
{\cal O}^{(8,1)}_{11}=\textrm{tr}[\lambda_{6} L_{\mu} S L^{\mu}],
& {\cal O}^{(27,1)}_{20}=t^{ij}_{kl}
(L_{\mu})^{k}_{i}(\partial_{\nu}W^{\mu
\nu})^{l}_{j},\\
{\cal O}^{(8,1)}_{12}=\textrm{tr}[\lambda_{6} L_{\mu}]
\textrm{tr}[\{L^{\mu},S\}], & {\cal O}^{(27,1)}_{24}=t^{ij}_{kl}
(W_{\mu \nu})^{k}_{i}(W^{\mu
 \nu})^{l}_{j},\\
{\cal
O}^{(8,1)}_{13}=\textrm{tr}[\lambda_{6} S] [L^{2}],\\
{\cal
O}^{(8,1)}_{15}=\textrm{tr}[\lambda_{6} [P,L^{2}]],\\
{\cal
O}^{(8,1)}_{35}=\textrm{tr}[\lambda_{6}\{L_{\mu},\partial_{\nu}W^{\mu
\nu}\}], \qquad\qquad\qquad\\
{\cal O}^{(8,1)}_{39}=\textrm{tr}[\lambda_{6} W_{\mu \nu} W^{\mu
\nu}],
\end{array}
\end{equation}
\medskip

\noindent with $ S=2B_{0}(\chi^{\dag}\Sigma + \Sigma^{\dag}\chi$),
$P=2B_{0}(\chi^{\dag}\Sigma-\Sigma^{\dag}\chi$), $L_{\mu}=i
\Sigma^{\dag}\partial_{\mu}\Sigma$ , and $W^{\mu
\nu}=2(\partial_{\mu}L_{\nu}+\partial_{\nu}L_{\mu})$.

This list is identical to that of Bijnens et al. \cite{bijnens}
except for the inclusion of $ {\cal O}^{(8,1)}_{35,39} $ and $
{\cal O}^{(27,1)}_{20,24} $ which contain surface terms, and so
cannot be absorbed into the other constants for processes which do
not conserve 4-momentum at the weak vertex.  Since we must use
4-momentum insertion in a number of our amplitudes, these
counterterms must be considered, and they are left explicit even
in the physical amplitudes.  There are additional operators
containing surface terms, but it was checked that these
counterterms can be absorbed into linear combinations of the above
minimal set for all amplitudes considered in this paper.

The $ \Delta S=2 $ operators are components of the same
irreducible tensor \cite{dono} under $ \textrm{SU}(3)_{L}\times
\textrm{SU}(3)_{R} $, and so the $ d_{i} $ are the same for both
the $ \Delta S=1 $ and $ \Delta S=2 $ cases.  The operators
governing $ \Delta S=2 $ transitions are obtained from the above $
(27_{L},1_{R})$'s, only with $ t^{33}_{22}=t^{22}_{33}=1,
t^{ij}_{kl}=0 $ otherwise.  This is important since some of our
information comes from the $ K^{0}\rightarrow \overline{K^{0}} $
amplitude.

The divergences associated with the counterterms have been
obtained in \cite{kambor} and \cite{bijnens}.  The subtraction
procedure can be defined as

\begin{equation}\label{8}
    e_{i}=e^{r}_{i}+\frac{1}{16
    \pi^{2}f^{2}}\left[\frac{1}{d-4}+\frac{1}{2}(\gamma_{E}-1-\ln 4
    \pi)\right]2(\alpha_{1}\varepsilon_{i}+\alpha_{2}\varepsilon'_{i}),
\end{equation}

\begin{equation}\label{9}
    d_{i}=d^{r}_{i}+\frac{1}{16
    \pi^{2}f^{2}}\left[\frac{1}{d-4}+\frac{1}{2}(\gamma_{E}-1-\ln 4
    \pi)\right]2 \alpha_{27} \gamma_{i},
\end{equation}

\noindent with the divergent pieces, $ \varepsilon_{i},
\varepsilon'_{i}, \gamma_{i} $ given in Table 1.

It is also necessary for the method of this paper to consider the
$O(p^{4})$ strong Lagrangian, which was first given by Gasser and
Leutwyler, ${\cal L}^{(4)}_{st}=\sum L_{i}{\cal O}^{(st)}_{i}$.

\begin{table}[htbp]
\caption{The divergences in the weak $O(p^4)$ counterterms,
$e_{i}$'s and $d_{i}$'s, for the (8,1)'s and (27,1)'s,
respectively. \label{tabone}}
\begin{center}
\begin{tabular}{|c|c|c|c|c|}
  \hline
   $ e_{i}$ & $ \varepsilon_{i}$ & $ \varepsilon'_{i} $ & $ d_{i} $ & $
   \gamma_{i} $  \\
  \hline
  1 & $ 1/4 $ & $ 5/6 $ &  1 & $ -1/6 $ \\
  2 & $ -13/18 $ & $ 11/18 $ &   2 & 0 \\
  3 & $ 5/12 $ & 0 &   4 & 3 \\
  4 & $ -5/36 $ & 0 &   5 & 1 \\
  5 & 0 & $ 5/12 $ &  6 & $ -3/2 $ \\
  10 & $ 19/24 $ & $ 3/4 $ &   7 & 1  \\
  11 & $ 3/4 $ & 0 &  20 & $ 1/2 $ \\
  12 & $ 1/8 $ & 0 &  24 & $ 1/8 $ \\
  13 & $ -7/8 $ & $ 1/2 $ & &  \\
  15 & $ 23/24 $ & $ -3/4 $ &  & \\
  35 & $ -3/8 $ & 0 & & \\
  39 & $ -3/16 $ & 0 & &  \\
  \hline
\end{tabular}
\end{center}
\end{table}

\noindent The strong $O(p^{4})$ operators relevant for
this calculation are the following \cite{gass}:\\

  \noindent \be \begin{array}{ll}{\cal
O}^{(st)}_{1}=\textrm{tr}[L^{2}]^{2},\\
    {\cal
O}^{(st)}_{2}=\textrm{tr}[L_{\mu}L_{\nu}]\textrm{tr}[L^{\mu}L^{\nu}],\\
     {\cal
O}^{(st)}_{3}=\textrm{tr}[L^{2}L^{2}],\\
     {\cal
O}^{(st)}_{4}=\textrm{tr}[L^{2}]\textrm{tr}[S],\\
     {\cal
O}^{(st)}_{5}=\textrm{tr}[L^{2}S],\\
     {\cal
O}^{(st)}_{6}=\textrm{tr}[S]^{2},\\
    {\cal
O}^{(st)}_{8}=\frac{1}{2} \textrm{tr}[S^{2}-P^{2}].\end{array} \ee\\

    The Gasser-Leuytwyler counterterms also contribute to the
    cancellation of divergencies in the expressions relevant to
    this paper.  The subtraction is defined similarly to that of
    the weak counterterms,

\begin{equation}\label{9}
    L_{i}=L^{r}_{i}+\frac{1}{16
    \pi^{2}}\left[\frac{1}{d-4}+\frac{1}{2}(\gamma_{E}-1-\ln 4
    \pi)\right] \Gamma_{i},
\end{equation}

\noindent with the divergent parts of the counterterm coefficients
given in Table 2  \cite{gass}.\\

\begin{table}[htbp]
\caption{The divergences in the strong $O(p^{4})$ counterterms,
$\Gamma_{i}$\cite{gass}. \label{tabtwo}}
\begin{center}
\begin{tabular}{|c|c|}
  \hline
  i & $\Gamma_{i}$ \\
  \hline
  1 & 3/32 \\
  2 & 3/16 \\
  3 & 0 \\
  4 & 1/8 \\
  5 & 3/8 \\
  6 & 11/144 \\
  8 & 5/48 \\
  \hline
\end{tabular}
\end{center}
\end{table}

\section{$ K \rightarrow \pi \pi $ amplitudes at $ O(p^{4}) $}

As mentioned before, in this work we will include both $ K
\rightarrow \pi $ with momentum insertion and $ K \rightarrow \pi
\pi $ at the two unphysical kinematics.  The complete list of
necessary ingredients consists of:  the two point functions $
K^{0} \rightarrow | 0 \rangle $, the three point functions $
K^{0}\rightarrow \overline{K^{0}}$, and $ K \rightarrow \pi $, all
with $ m_{s}\neq m_{d}=m_{u} $, and the four  point functions $ K
\rightarrow \pi \pi $ at the two values of unphysical kinematics,
$ m_{K}=m_{\pi} $ (requiring energy insertion), and $
m_{K}=2m_{\pi} $.  These two threshold values of the kinematics
bypass the Miani-Testa theorem, which states that multihadron
final states are not accessible on the lattice at any other
kinematics aside from the threshold \cite{maiani}. At these
kinematics, the strong phases are 0, and the effects of final
state interactions vanish.  However, these amplitudes at
unphysical kinematics do contain information on the $ O(p^{4}) $
low energy constants, and when combined with information from the
other two- and three-point functions mentioned above, all of the $
O(p^{4}) $ low energy constants necessary for $ K \rightarrow \pi
\pi $ can be obtained. The phases of the amplitude are introduced
in ChPT via the one-loop unitarity corrections of the $ O(p^{2}) $
operators.

Because the $ K \rightarrow \pi $ amplitudes do not conserve
four-momentum for $ m_{s}\neq m_{d} $, it is necessary to allow
the weak operator to transfer a four-momentum $ q \equiv
p_K-p_\pi$, as in \cite{cirig}.  This is also necessary for the
case of $ K \rightarrow \pi \pi , m_{K}=m_{\pi} $ \cite{berntwo}.
At $ O(p^{4}) $, this requires the inclusion of (potentially many)
surface terms in our minimal counterterm operator basis.  The
number of additional such terms appearing in linearly independent
combinations was discovered to be small (four), and thus the
method was not invalidated.  Also, this method requires the
computation of $ K \rightarrow \pi \pi $ matrix elements at
unphysical kinematics because there are LEC's which appear in $ K
\rightarrow \pi \pi $ but do not appear in $ K \rightarrow \pi $
at all.  These are $ d_{5}, e_{13} $ and $ e_{15} $.

The diagrams which must be evaluated for the $O(p^{4})$
corrections are shown in Figure 1.  The diagrams to be evaluated
for $K\to |0\rangle$ are A1 and A2.  The diagrams to be evaluated
for $K\to\pi$ and $K\to\overline{K}$ are B1-B3. C1-C6 and D1-D6
must be evaluated for $K\to\pi \pi$.  D1-D6 contain the tadpole
vertex of the weak mass $O(p^{2})$, (8,1) operator.  Also, the
renormalization of the external legs via the strong interaction
must be taken into account.

\begin{figure}[htbp]
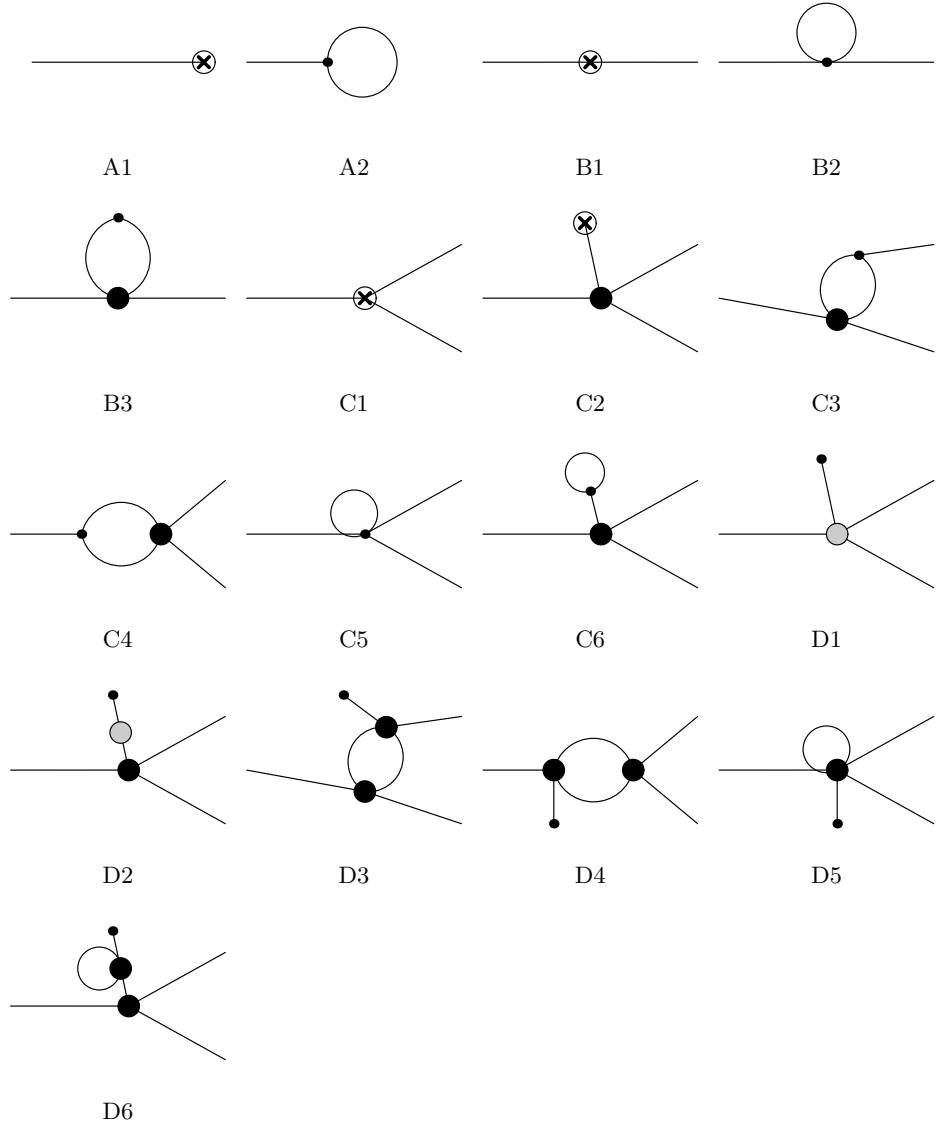


\unitlength=1bp%

\begin{feynartspicture}(432,445)(4,5)

\FADiagram{A1} \FAProp(2.,10.)(18.,10.)(0.,){/Straight}{0}
\FAVert(18.,10.){2}

\FADiagram{A2} \FAProp(0.,10.)(7.5,10.)(0.,){/Straight}{0}
\FAProp(7.5,10.)(7.5,10.)(14.,10.){/Straight}{0}
\FAVert(7.5,10.){0}

\FADiagram{B1} \FAProp(0.,10.)(10.,10.)(0.,){/Straight}{0}
\FAProp(20.,10.)(10.,10.)(0.,){/Straight}{0} \FAVert(10.,10.){2}

\FADiagram{B2} \FAProp(0.,10.)(10.,10.)(0.,){/Straight}{0}
\FAProp(20.,10.)(10.,10.)(0.,){/Straight}{0}
\FAProp(10.,10.)(10.,10.)(10.,15.5){/Straight}{0}
\FAVert(10.,10.){0}

\FADiagram{B3} \FAProp(0.,10.)(10.,10.)(0.,){/Straight}{0}
\FAProp(20.,10.)(10.,10.)(0.,){/Straight}{0}
\FAProp(10.,17.5)(10.,10.)(-0.8,){/Straight}{0}
\FAProp(10.,17.5)(10.,10.)(0.8,){/Straight}{0}
\FAVert(10.,10.){-5} \FAVert(10.,17.5){0}

\FADiagram{C1} \FAProp(0.,10.)(11.,10.)(0.,){/Straight}{0}
\FAProp(20.,15.)(11.,10.)(0.,){/Straight}{0}
\FAProp(20.,5.)(11.,10.)(0.,){/Straight}{0} \FAVert(11.,10.){2}

\FADiagram{C2} \FAProp(0.,10.)(11.,10.)(0.,){/Straight}{0}
\FAProp(20.,15.)(11.,10.)(0.,){/Straight}{0}
\FAProp(20.,5.)(11.,10.)(0.,){/Straight}{0}
\FAProp(9.5,17.)(11.,10.)(0.,){/Straight}{0} \FAVert(11.,10.){-5}
\FAVert(9.5,17.){2}

\FADiagram{C3} \FAProp(0.,10.)(11.,8.)(0.,){/Straight}{0}
\FAProp(20.,15.)(13.,14.)(0.,){/Straight}{0}
\FAProp(20.,5.)(11.,8.)(0.,){/Straight}{0}
\FAProp(13.,14.)(11.,8.)(0.8,){/Straight}{0}
\FAProp(13.,14.)(11.,8.)(-0.8,){/Straight}{0} \FAVert(13.,14.){0}
\FAVert(11.,8.){-5}

\FADiagram{C4} \FAProp(0.,10.)(6.6,10.)(0.,){/Straight}{0}
\FAProp(20.,15.)(14.,10.)(0.,){/Straight}{0}
\FAProp(20.,5.)(14.,10.)(0.,){/Straight}{0}
\FAProp(6.6,10.)(14.,10.)(0.8,){/Straight}{0}
\FAProp(6.6,10.)(14.,10.)(-0.8,){/Straight}{0} \FAVert(6.6,10.){0}
\FAVert(14.,10.){-5}

\FADiagram{C5} \FAProp(0,10.)(11.,10.)(0,){/Straight}{0}
\FAProp(20.,15.)(11.,10.)(0,){/Straight}{0}
\FAProp(20.,5.)(11.,10.)(0,){/Straight}{0}
\FAProp(11.,10.)(11.,10.)(9.0034,13.8721){/Straight}{0}
\FAVert(11.,10.){0}

\FADiagram{C6} \FAProp(0.,10.)(11.,10.)(0.,){/Straight}{0}
\FAProp(20.,15.)(11.,10.)(0.,){/Straight}{0}
\FAProp(20.,5.)(11.,10.)(0.,){/Straight}{0}
\FAProp(10.,14.)(11.,10.)(0.,){/Straight}{0}
\FAProp(10.,14.)(10.,14.)(9.,17.5){/Straight}{0}
\FAVert(11.,10.){-5} \FAVert(10.,14.){0}

\FADiagram{D1} \FAProp(0.,10.)(11.,10.)(0.,){/Straight}{0}
\FAProp(20.,15.)(11.,10.)(0.,){/Straight}{0}
\FAProp(20.,5.)(11.,10.)(0.,){/Straight}{0}
\FAProp(9.5,17.)(11.,10.)(0.,){/Straight}{0} \FAVert(11.,10.){-1}
\FAVert(9.5,17.){0}

\FADiagram{D2} \FAProp(0,10.)(11.,10.)(0,){/Straight}{0}
\FAProp(20.,15.)(11.,10.)(0,){/Straight}{0}
\FAProp(20.,5.)(11.,10.)(0,){/Straight}{0}
\FAProp(9.5,17)(10.25,13.5)(0,){/Straight}{0}
\FAProp(10.25,13.5)(11.,10.)(0,){/Straight}{0}
\FAVert(11.,10.){-5} \FAVert(9.5,17.){0} \FAVert(10.25,13.5){-1}

\FADiagram{D3} \FAProp(0.,10.)(11.,8.)(0.,){/Straight}{0}
\FAProp(20.,15.)(13.,14.)(0.,){/Straight}{0}
\FAProp(20.,5.)(11.,8.)(0.,){/Straight}{0}
\FAProp(13.,14.)(11.,8.)(0.8,){/Straight}{0}
\FAProp(13.,14.)(11.,8.)(-0.8,){/Straight}{0}
\FAProp(13.,14.)(9.,17.)(0.,){/Straight}{0} \FAVert(13.,14.){-5}
\FAVert(11.,8.){-5} \FAVert(9.,17.){0}

\FADiagram{D4} \FAProp(0.,10.)(6.6,10.)(0.,){/Straight}{0}
\FAProp(20.,15.)(14.,10.)(0.,){/Straight}{0}
\FAProp(20.,5.)(14.,10.)(0.,){/Straight}{0}
\FAProp(6.6,10.)(14.,10.)(0.8,){/Straight}{0}
\FAProp(6.6,10.)(14.,10.)(-0.8,){/Straight}{0}
\FAProp(6.6,10.)(6.6,5.)(0.,){/Straight}{0} \FAVert(6.6,10.){-5}
\FAVert(14.,10.){-5} \FAVert(6.6,5.){0}

\FADiagram{D5} \FAProp(0,10.)(11.,10.)(0,){/Straight}{0}
\FAProp(20.,15.)(11.,10.)(0,){/Straight}{0}
\FAProp(20.,5.)(11.,10.)(0,){/Straight}{0}
\FAProp(11.,10.)(11.,10.)(9.0034,13.8721){/Straight}{0}
\FAProp(11.,10.)(11.,5.)(0.,){/Straight}{0} \FAVert(11.,10.){-5}
\FAVert(11.,5.){0}

\FADiagram{D6} \FAProp(0,10.)(11.,10.)(0,){/Straight}{0}
\FAProp(20.,15.)(11.,10.)(0,){/Straight}{0}
\FAProp(20.,5.)(11.,10.)(0,){/Straight}{0}
\FAProp(9.5,17)(10.25,13.5)(0,){/Straight}{0}
\FAProp(10.25,13.5)(11.,10.)(0,){/Straight}{0}
\FAProp(10.25,13.5)(10.25,13.5)(6.25,13.5){/Straight}{0}
\FAVert(11.,10.){-5} \FAVert(9.5,17.){0} \FAVert(10.25,13.5){-5}

\end{feynartspicture}

\caption{$O(p^{4})$ corrections include tree-level diagrams with
insertion of the $O(p^{4})$ weak vertices (crossed circles),
tree-level diagrams with insertion of $O(p^{4})$ strong vertices
(lightly shaded circles), one-loop diagrams with insertions of the
$O(p^{2})$ weak vertices (small filled circles) and the $O(p^{2})$
strong vertices (big filled circles).  A1 and A2 are for $K\to
|0\rangle$. B1-B3 are for $K\to\pi$ and $K\to\overline{K}$.  C1-C6
and D1-D6 are for $K\to\pi \pi$.}
\end{figure}

\subsection{$ (27_{L},1_{R})$, $ \Delta I=3/2 $}

The counterterms necessary to reconstruct the $ O(p^{4})
(27_{L},1_{R}), \Delta I=3/2, K \rightarrow \pi \pi $ amplitudes
can be obtained from $ K^{0}\rightarrow \overline{K^{0}}$; $ K^{+}
\rightarrow \pi^{+}, \Delta I=3/2 $; and $ K \rightarrow \pi \pi,
\Delta I=3/2 $ at {\it either} value of the unphysical kinematics.
The expression for $ K^{0}\rightarrow \overline{K^{0}} $ is given
by (all masses and decay constants are the bare ones)\\


\begin{eqnarray}\label{10}
    \langle \overline{K^0}|{\cal
O}^{(27,1)}_{\Delta
    S=2}|K^{0}\rangle_{ct}& =
    &\frac{8\alpha_{27}}{f^{2}}m^{2}_{K}-\frac{8}{f^{2}}
    [4(d^{r}_{1}+d^{r}_{2}+d^{r}_{20}-4d^{r}_{24}-d^{r}_{4}-d^{r}_{7})m^{4}_{K}\nonumber \\
        & &
        -2(4d^{r}_{1}+d^{r}_{7})m^{2}_{K}m^{2}_{\pi}+4d^{r}_{1}m^{4}_{\pi}]
\end{eqnarray}

Equation~(21), as well as all the following amplitudes, include
only the tree level $ O(p^{2}) $ and $ O(p^{4}) $ weak counterterm
contributions.  For brevity, the logarithmic terms, as well as the
Gasser-Leutwyler $ L_{i} $ counterterms have been omitted in the
body of the paper, but are included in a set of Appendixes.  It
was verified that the divergences in the logarithmic terms cancel
those of the counterterms, providing a strong check on the
calculation.  Note also that for the application of this method
most of the Gasser-Leutwyler counterterms must be known, and that
an improved determination of the relevant ones could be obtained
from a lattice calculation of observables in the purely strong
sector, e.g. most can be obtained from the pseudoscalar masses and
decay constants.

From the above $ K^{0}\rightarrow \overline{K^{0}}$ amplitude, one
can extract the values of $ d^{r}_{1} $ and $ d^{r}_{7} $ from a
fit to terms quadratic in the quark masses.  The other relevant
expressions for $ K \rightarrow \pi \pi , \Delta I=3/2 $ are

\begin{eqnarray}\label{11}
&&\langle \pi^{+}|{\cal O}^{(27,1),(3/2)}|K^{+}\rangle_{ct} =
-\frac{4}{f^{2}} \alpha_{27}p_{K}\cdot p_{\pi}
 +\frac{8}{f^{2}}[2d^{r}_{2}m^{2}_{K}m^{2}_{\pi}  +(d^{r}_{20}
 -d^{r}_{4} -2d^{r}_{7}) \nonumber \\
&&\qquad\qquad m^{2}_{K}p_{K}\cdot
 p_{\pi} +(d^{r}_{20}-d^{r}_{4}-d^{r}_{7})m^{2}_{\pi}p_{K}\cdot
p_{\pi} - 8d^{r}_{24}(p_{K}\cdot p_{\pi})^{2}],
\end{eqnarray}

\begin{equation}\label{12}
\langle \pi^{+} \pi^{-}|{\cal
O}^{(27,1),(3/2)}|K^{0}\rangle_{ct}=-\frac{8i\alpha_{27}}{f^{3}}
m^{2} +
\frac{16im^{4}}{f^{3}}(d^{r}_{20}-2d^{r}_{4}+d^{r}_{5}-3d^{r}_{7}),
\end{equation}

\noindent for $ K \rightarrow \pi \pi, m_{K}=m_{\pi}=m, $ and

\begin{equation}
\langle \pi^{+} \pi^{-}|{\cal O}^{(27,1),(3/2)}|K^{0}\rangle_{ct}
= -\frac{4i\alpha_{27}}{f^{3}}(m^{2}_{K}-m^{2}_{\pi})
+\frac{3im^{4}_{K}}{2f^{3}}              
(2d^{r}_{2}+2d^{r}_{20}
-8d^{r}_{24}-4d^{r}_{4}+2d^{r}_{5}-9d^{r}_{7}),
\end{equation}

\noindent for $ K \rightarrow \pi \pi,
m_{K(1-loop)}=2m_{\pi(1-loop)}. $

From Eq.~(22) we get the additional combinations of counterterms $
d^{r}_{2}, d^{r}_{24}, $ and $ d^{r}_{4}-d^{r}_{20} $.  From
either expression for $ K \rightarrow \pi \pi $ at unphysical
kinematics we can then obtain $ d^{r}_{4}-d^{r}_{5} $.  Along with
the tree level LEC, $\alpha_{27} $, these five linear combinations
$[d^r_2,d^r_7, d^r_4-d^r_5, d^r_4-d^r_{20}, d^r_{24}]$  are
sufficient to determine $ K \rightarrow \pi \pi $ at the physical
kinematics as given in Eq. (25).  Notice that there is
considerable redundancy in determining these coefficients. For
example, $d_4-d_{20}$, $d_4-d_5$, $d_2\dots$ occur in several of
Eqs. (21)--(24)\cite{spqcdr2}.

\begin{eqnarray}\label{14}
&&\langle \pi^{+} \pi^{-}|{\cal
O}^{(27,1),(3/2)}|K^{0}\rangle_{ct} = -\frac{4i\alpha_{27}}{f_K
f^{2}_\pi}(m^{2}_{K}-m^{2}_{\pi})_{(1-loop)} +
\frac{4i}{f_K f^{2}_\pi}(m^{2}_{K}-m^{2}_{\pi})\nonumber \\
&&[(-d^{r}_{4}+d_{5}^{r}
-4d^{r}_{7})m^{2}_{K}
 +(4d^{r}_{2}+4d^{r}_{20}-16d^{r}_{24}-4d^{r}_{4}-2d^{r}_{7})m^{2}_{\pi}].
\end{eqnarray}

The logarithmic and Gasser-Leutwyler counterterm contributions to
the expressions in this subsection are given in Appendix C.  Note,
also, that for the cases of physical $K\to\pi \pi$ amplitudes,
Eqs. (25),(34), and (35), the pseudoscalar decay constants and
masses are the physical (renormalized to one-loop order) ones. For
all other amplitudes given in this paper except $K\to\pi \pi$ at
physical kinematics the formulas are in terms of the bare
constants.  The distinction between bare and renormalized
constants is made only in tree-level amplitudes, since making this
distinction in the $O(p^4)$ expressions introduces corrections at
higher order [$O(p^6)$] than is considered in this paper.

\subsection{$ (8_{L},1_{R})+(27_{L},1_{R})$, $ \Delta I=1/2 $}

The counterterms necessary to reconstruct the $ O(p^{4})$
$[(8_{L},1_{R})+(27_{L},1_{R})]$, $ \Delta I=1/2, K \rightarrow
\pi \pi $ amplitudes, relevant for operators such as $Q^{1/2}_2$,
can be obtained using $d_i$'s obtained from the $[(27,1);\Delta
I=3/2]$ case given above along with information from $ K^{0}
\rightarrow |0\rangle; K^{+}\rightarrow \pi^{+}, \Delta I=1/2 $;
and $ K \rightarrow \pi \pi, \Delta I=1/2 $ at {\it both\/}
unphysical kinematics. For $ K^{0} \rightarrow |0\rangle $, we
have

\begin{eqnarray}\label{15}
&&\langle 0|{\cal O}^{(8,1)}|K^{0}\rangle_{ct} =
\frac{4i\alpha_{2}}{f}(m^{2}_{K}-m^{2}_{\pi})
 -
\frac{8i}{f}[2(-e^{r}_{1}-e^{r}_{2}+e^{r}_{5})m^{4}_{K}\nonumber \\
&& + (2e^{r}_{1}+e^{r}_{2} -2e^{r}_{5})m^{2}_{K}m^{2}_{\pi}
+ e^{r}_{2}m^{4}_{\pi}],
\end{eqnarray}

\begin{equation}\label{16}
\langle 0|{\cal
O}^{(27,1)}|K^{0}\rangle_{ct}=-\frac{48i}{f}d^{r}_{1}(m^{2}_{K}-m^{2}_{\pi}
)^{2}.
\end{equation}

Given the previously obtained value of $ d^{r}_{1} $ from the $
\Delta I=\frac32 $ case, we can obtain $ e^{r}_{2} $ and $
e^{r}_{1}-e^{r}_{5} $ from $ K^{0} \rightarrow |0\rangle $.  The
other relevant expressions are

\begin{eqnarray}\label{17}
&&\langle \pi^{+}|{\cal O}^{(27,1),(1/2)}|K^{+}\rangle_{ct} =
-\frac{4}{f^{2}}\alpha_{27}p_{K}\cdot p_{\pi}  -
\frac{8}{f^{2}}[6d^{r}_{1}m^{4}_{K}-2(3d^{r}_{1}+d^{r}_{2})m^{2}_{K}
m^{2}_{\pi} \nonumber \\
&&+ (-d^{r}_{20}+d^{r}_{4}-3d^{r}_{6}+2d^{r}_{7})m^{2}_{K}p_{K}\cdot
p_{\pi}  +
(-d^{r}_{20}+d^{r}_{4}+3d^{r}_{6}+d^{r}_{7})m^{2}_{\pi}p_{K}\cdot
p_{\pi}\nonumber \\
&&\qquad\qquad\qquad\qquad\qquad+8d^{r}_{24}(p_{K}\cdot p_{\pi})^{2}],
\end{eqnarray}

\begin{eqnarray}\label{18}
&&\langle \pi^{+}|{\cal O}^{(8,1),(1/2)}|K^{+}\rangle_{ct} =
\frac{4}{f^{2}}\alpha_{1}p_{K}\cdot p_{\pi} -
\frac{4}{f^{2}}\alpha_{2}m^{2}_{K}  -
\frac{8}{f^{2}}[2(e^{r}_{1}+e^{r}_{2}-e^{r}_{5})m^{4}_{K}\nonumber \\
&&+ (e^{r}_{2}+2e^{r}_{3} + 2e^{r}_{5})
m^{2}_{K}m^{2}_{\pi}
 + (2e^{r}_{35}-2e^{r}_{10})m^{2}_{K}p_{K}\cdot p_{\pi} +
(2e^{r}_{35}-e^{r}_{11})\nonumber \\
&&\qquad\qquad\qquad m^{2}_{\pi}p_{K}\cdot
p_{\pi}-8e^{r}_{39}(p_{K}\cdot p_{\pi})^{2}],
\end{eqnarray}

\noindent as well as

\begin{equation}\label{19}
\langle \pi^{+} \pi^{-}|{\cal
O}^{(27,1),(1/2)}|K^{0}\rangle_{ct}=-8i\frac{\alpha_{27}
}{f^{3}}m^{2} +
16i\frac{m^{4}}{f^{3}}(d^{r}_{20}-2d^{r}_{4}+d^{r}_{5}-3d^{
r}_{7}),
\end{equation}

\begin{eqnarray}\label{20}
&&\langle \pi^{+} \pi^{-}|{\cal O}^{(8,1),(1/2)}|K^{0}\rangle_{ct}
= 8i \frac{\alpha_{1}}{f^{3}}m^{2}+4i\frac{\alpha_{2}}{f^{3}}m^{2}
  + 8i\frac{m^{4}}{f^{3}}\nonumber \\
&&(2e^{r}_{1}+4e^{r}_{10}+ 2e^{r}_{11}+4e^{r}_{15}
 + 3e^{r}_{2}-4e^{r}_{35}-2e^{r}_{5}),
\end{eqnarray}

\noindent for $ K \rightarrow \pi \pi, m_{K}=m_{\pi}=m, $ and

\begin{eqnarray}\label{21}
&&\langle \pi^{+} \pi^{-}|{\cal
O}^{(27,1),(1/2)}|K^{0}\rangle_{ct} =
-4i\frac{\alpha_{27}}{f^{3}}(m^{2}_{K}-m^2_\pi)
 + \frac{3i}{2}\frac{m^{4}_{K}}{f^{3}}\nonumber \\
 &&(6d^{r}_{1}+
2d^{r}_{2}+2d^{r}_{20}-8d^{r}_{24} -4d^{r}_{4}+2d^{r}_{5}+12d^{r}_{6}-9d^{r}
_{7}),
\end{eqnarray}

\begin{eqnarray}\label{22}
&&\langle \pi^{+} \pi^{-}|{\cal O}^{(8,1),(1/2)}|K^{0}\rangle_{ct}
= 4i\frac{\alpha_{1}}{f^{3}}(m^{2}_{K}-m^2_\pi)
 + \frac{3i}{2}\frac{m^{4}_{K}}{f^{3}}\nonumber \\
 && (-2e^{r}_{1}+
6e^{r}_{10}+e^{r}_{11}-4e^{r}_{13}
 + 4e^{r}_{15}-4e^{r}_{2}-2e^{r}_{3}-4e^{r}_{35}+8e^{r}_{39})
\end{eqnarray}

\noindent for $ K \rightarrow \pi \pi,
m_{K(1-loop)}=2m_{\pi(1-loop)}. $

From expressions (28) and (29) one can obtain the leading order
LEC's $ \alpha_{1},$ and $ \alpha_{2}$, as well as the linear
combinations $ e^{r}_{39}, e^{r}_{1}+e^{r}_{3},
e^{r}_{10}-e^{r}_{35}+\frac{3}{2}d^{r}_{6}, $ and $
2e^{r}_{10}-e^{r}_{11}+6d^{r}_{6}. $  From Eqs.~(30) and (31) for
UK1 one can then obtain $ e^{r}_{11}+2e^{r}_{15}-3d^{r}_{6}.$
Making use of all the input thus obtained into Eqs.~(32,33) for
UK2 yields $e^r_{13}-\frac{3}{2}d^r_6$. These 14 linear
combinations (namely $d^r_1, d^r_2, d^r_7, d^r_4-d^r_5,
d^r_4-d^r_{20}, d^r_{24}, e^r_2, e^r_1-e^r_5, e^r_1+e^r_3,
e^r_{39}, e^r_{10}-e^r_{35} +\frac32 d^r_6,
2e^r_{10}-e^r_{11}+6d^r_6, e^r_{11}+2e^r_{15} -3 d^r_6,
e^r_{13}-\frac32d^r_6$) are sufficient to reconstruct the physical
$ K \rightarrow \pi \pi, \Delta I=1/2 $ amplitudes for operators
such as $Q^{1/2}_1, Q^{1/2}_2$ etc.:

\begin{eqnarray}\label{24}
&& \langle \pi^{+} \pi^{-}|{\cal
O}^{(27,1),(1/2)}|K^{0}\rangle_{ct} = -4i\frac{\alpha_{27}}{f_K
f^{2}_\pi}(m^{2}_{K}-m^{2}_{\pi})_{(1-loop)}
 + 4i\frac{1}{f_K f^{2}_\pi}(m^{2}_{K}-m^{2}_{\pi})\nonumber \\
&&\qquad\quad[(-d^{r}_{4}+d^{r}_{5}+9d^{r}_{6}-4d^{r}_{7})m^{2}_{K}
 + 2(6d^{r}_{1}+2d^{r}_{2}+2d^{r}_{20}-8d^{r}_{24} \nonumber \\
 &&\qquad\qquad\qquad\qquad -2d^{r}_{4}
-6d^{r}_{6}-d^{r}_{7})m^{2}_{\pi}],
\end{eqnarray}

\begin{eqnarray}\label{25}
&&\langle \pi^{+} \pi^{-}|{\cal O}^{(8,1),(1/2)}|K^{0}\rangle_{ct}
= 4i\frac{\alpha_{1}}{f_K
f^{2}_\pi}(m^{2}_{K}-m^{2}_{\pi})_{(1-loop)}
  +
  8i\frac{1}{f_K f^{2}_\pi}(m^{2}_{K}-m^{2}_{\pi})\nonumber \\
&&\qquad\quad[(e^{r}_{10}-2e^{r}_{13}+e^{r}_{15})m^{2}_{K}
 + (-2e^{r}_{1}+2e^{r}_{10}+e^{r}_{11}+4e^{r}_{13}-4e^{r}_{2}\nonumber \\
&&\qquad\qquad\qquad\qquad -2e^{r}_{3}-4e^{r}_{35}+8e^{r}_{39})m^{2}_{\pi}].
\end{eqnarray}\\

The logarithmic and Gasser-Leutwyler counterterm contributions to
the amplitudes presented in this subsection are given in Appendix
D.

\subsection{$(8_L,1_R)$, $\Delta I=1/2$}

The case of pure (8,1) operators, i.e. $Q_{3,4,5,6}$ is simpler
than the previous case of mixed $\Delta I=1/2$ operators; note
also that phenomenologically, pure (8,1)'s are the most important.
This is clearly a special case of the previous one for which
$(27_L,1_R)$ contributions are irrelevant. For the physical
$K\to\pi\pi$ reaction at $O(p^4)$, Eq.~(35), eight new linear
combinations are needed: $e^r_2, e^r_1-e^r_5, e^r_3+e^r_5,
e^r_{35}-e^r_{10}, 2e^r_{35}-e^r_{11}, e^r_{39},
e^r_{11}+2e^r_{15}, e^r_{13}$.

The terms quadratic in quark mass for $K\to0$, Eq.~(26), yield
$e^r_2$ and $e^r_1$--$e^r_5$. A similar fit to $K^+\to\pi^+$,
Eq.~(29), then leads to $e^r_3+e^r_5, e^r_{35}-e^r_{10},
2e^r_{35}-e^r_{11}$ and $e^r_{39}$. Using this for $K\to\pi\pi$ at
UK1, Eq.~(31), yields $e^r_{11}+2e^r_{15}$ and $K\to\pi\pi$ at
UK2, Eq.~(33), may be fitted to give $e^r_{13}$. While determining
these coefficients is expected to be quite demanding, it is useful
to note that several of them are obtained via more than one
measurement. Note, in particular, that the term linear in quark
mass, $\alpha_2$, originating from operator mixing occurs in
$K\to0$, in $K\to\pi$ and also in $K\to\pi\pi$ at UK1 where the
operator injects energy.

The logarithmic and Gasser-Leutwyler counterterm contributions to
the amplitudes presented in this subsection are a subset
of those given in Appendix
D.

\section{Conclusion}

This paper presents all of the counterterm and finite logarithm
contributions to $ K^{0} \rightarrow | 0 \rangle $, $
K^{0}\rightarrow \overline{K^{0}}$, and $ K \rightarrow \pi $ with
momentum insertion, and $ K \rightarrow \pi \pi $ (at two values
of unphysical kinematics) to $ O(p^{4}) $ in ChPT for the $
(27_{L},1_{R}) $ and $ (8_{L},1_{R}) $ operators.  It demonstrates
that these quantities are sufficient to fully determine $ K
\rightarrow \pi \pi $ to $O(p^4)$ at the physical kinematics.  It
should be emphasized that this calculation was done in full ChPT,
and that these arguments do not necessarily apply to the quenched
theory. In fact it is quite likely that some of the $K\to\pi\pi$
matrix elements suffer from large corrections due to the quenched
approximation; this possibilty has recently been raised in Ref.\
\cite{golttwo} for the case of $Q_6$. Indeed, we have done a fit
to the quenched RBC data \cite{blum} for $ Q_{7}^{3/2} $ and $
Q_{8}^{3/2} $ using the next-to-leading order ChPT prediction of
Cirigliano and Golowich \cite{cirig} and have found a poor fit ($
\chi^{2}/d.o.f. \approx 2 $).  Thus, the data tends to disfavor a
large coefficient for the chiral log term that is predicted by
full ChPT\null.   A simple quadratic fit with the coefficient of
the log term set to 0 yielded a much better fit
$(\chi^{2}/d.o.f.\approx 0.1)$.  These arguments suggest that an
unquenched lattice calculation is probably necessary in order to
correctly extract the $ O(p^{4}) $ counterterms from the lattice.
It is clearly important to see whether this extraction procedure,
especially including $K\to\pi\pi$ at the two unphysical
kinematics, can be extended to the case of $O(p^{4})$ quenched
ChPT.

In closing we briefly want to remind the reader that two other
interesting methods have been proposed recently \cite{lell,buch}
for lattice extraction of $K\to\pi\pi$ amplitudes. We believe it
is important to use all the methods in order to obtain reliable
information on this important process.

\bigskip
\bigskip
\bigskip

\centerline{\bf ACKNOWLEDGEMENTS}

This research was supported in part by US DOE Contract No.
DE-AC02-98CH10886. We thank Joachim Kambor and Carl Wolfe for
providing us copies of their Ph.D theses \cite{kamb,wolfe}, Tom
Blum for discussions and Maarten Golterman,
Guido Martinelli, Elisabetta Pallante and Chris Sachrajda for communications.

\bigskip
\bigskip
\bigskip

\appendix
\section*{\bf APPENDIX A}
\setcounter{equation}{0}
\setcounter{section}{1}
\renewcommand{\theequation}{A\arabic{equation}}

\bigskip

Appendixes B-D contain the finite logarithm and Gasser-Leutwyler
counterterm $O(p^{4})$ contributions to the amplitudes presented
in this paper.  They were calculated using the \textsc{FeynCalc}
package \cite{mert} written for the \textsc{Mathematica}
\cite{wolfram} system. These expressions involve the regularized
Veltman-Passarino basis integrals $A$ and $B$ \cite{pass}:

\be A(m^2) = \frac{1}{16\pi^2f^2} m^2 \ln \frac{m^2}{\mu^2}
=\lim_{d\to4}\frac1i\left[ \mu^{4-d}\int \frac{d^d\ell}{(2\pi)^d}
\; \frac{1}{\ell^2-m^2} + 2i m^2\bar\lambda \right], \ee

\bea B(q^2,m^2_1,m^2_2)\!\!\!\!\! & = &
\!\!\!\!\!\lim_{d\to4}\frac1i \left[ \mu^{4-d}\int
\frac{d^d\ell}{(2\pi)^d} \frac{1}{((\ell+q)^2 -m^2_1) (\ell^2
-m^2_2)} + 2i\bar\lambda\right]\nonumber \\
\!\!\!\!\!& = &\!\!\!\!\!
\int^1_0{dx} \frac{1}{(4\pi)^2} [1+\ln (-x(1-x) q^2 +xm^2_1
+(1-x)m^2_2)\nonumber \\
\!\!\!\!\!& &\!\!\!\!\! -\ln \mu^2 ], \eea

where

\be \bar\lambda = \frac{1}{16\pi^2} \left[ \frac{1}{d-4} -\frac12
(\ln4\pi-\gamma_E +1)\right]. \ee

\bigskip
\bigskip
\bigskip

\appendix
\section*{\bf APPENDIX B}
\setcounter{equation}{0} \setcounter{section}{1}
\renewcommand{\theequation}{B\arabic{equation}}
\bigskip

At 1-loop order, the pseudoscalar decay constants and masses are
renormalized such that $f_{\pi,K}=f\left(1+\frac{\Delta
f_{\pi,K}}{f}\right)$ and $m^2_{\pi,K
(1-loop)}=m^2_{\pi,K}\left(1+\frac{\Delta
m^2_{\pi,K}}{m^2_{\pi,K}}\right)$.  The corrections are given by

\bea \frac{\Delta f_\pi}{f}\!\!\!\!\!  & = &\!\!\!\!\!
-2A(m^2_\pi) - A(m^2_K) + \frac{8}{f^2}
(2m^2_K+m^2_\pi) L_4 + \frac{8}{f^2} m^2_\pi L_5, \\
\frac{\Delta f_K}{f}\!\!\!\!\! & = &\!\!\!\!\!-\frac34 A(m^2_\pi)
-\frac32 A(m^2_K) -\frac34 A(m^2_\eta)  +\frac{8}{f^2} (2m^2_K
+m^2_\pi) L_4 \nonumber \\ & &+\frac{8}{f^2} m^2_K L_5. \eea

\bea \frac{\Delta m^2_\pi}{m^2_\pi} & = &  A(m^2_\pi) -\frac13
A(m^2_\eta) +
\frac{16}{f^2} [(-L_4+2L_6) 2m^2_K \\
& & + (-L_4-L_5 +2L_6+2L_8) m^2_\pi ], \nonumber \\
\frac{\Delta m^2_K}{m^2_K}  & = &  \frac23 A(m^2_\eta) +
\frac{16}{f^2}
[(-2L_4-L_5+4L_6+2L_8) m^2_K \\
& & + (-L_4+2L_6) m^2_\pi].\nonumber \eea

\noindent For degenerate quark masses at 1-loop order, $m^{2}_{K
(1-loop)}=m^{2}_{\pi (1-loop)}=m^{2}\left(1+\frac{\Delta
m^2}{m^2}\right)$, $f_\pi=f_K=f\left(1+\frac{\Delta f}{f}\right)$,

\bea \frac{\Delta m^2}{m^2} & = & \frac23 A(m^2) +
\frac{16m^2}{f^2}
(-3L_4-L_5 +6L_6 + 2L_8), \\
\frac{\Delta f}{f} & = & -3A(m^2) + \frac{8m^2}{f^2} (3L_4 + L_5).
\eea

\bigskip
\bigskip

\appendix
\section*{\bf APPENDIX C}
\setcounter{equation}{0} \setcounter{section}{1}
\renewcommand{\theequation}{C\arabic{equation}}
\bigskip

The logarithmic corrections for the quantities relevant for the
determination of the (27,1), $\Delta I=3/2$ $K\to\pi\pi$
amplitudes in this paper are given by

\bea \bra \bar K^0|{\cal O}^{(27,1)}_{\Delta S=2}|K^0\ket_{log} &
= & \frac{8\alpha_{27}}{f^2} \left[ -\frac{2}{16\pi^2f^2} m^4_K
-8m^2_KA(m^2_K) + \left(\frac12 m^2_\pi -\frac{13}{2} m^2_K\right)
A(m^2_\eta)\right. \nonumber\\
& & \left. + \left(-\frac72 m^2_K -\frac12 m^2_\pi\right)
A(m^2_\pi) -2 \frac{\Delta f_K}{f} m^2_K  +\Delta m^2_K\right],
\eea

\bea \bra \pi^+ |{\cal O}^{(27,1),(3/2)} |K^+\ket_{log} & = &
-\frac{4\alpha_{27}}{f^2} p_K\cdot p_\pi \left[ -2p_K \cdot p_\pi
B(q^2,
m^2_K, m^2_\pi) - \frac32 A(m^2_\eta) - 7A(m^2_K) \right. \nonumber \\
& & \left.- \frac{15}{2} A(m^2_\pi)- \frac{\Delta f_K}{f} -
\frac{\Delta f_\pi}{f} \right], \eea

\be \bra\pi^+\pi^- | {\cal O}^{(27,1),(3/2)} | K^0\ket_{log} = -8i
\frac{\alpha_{27}}{f^3} m^2 \left[ \frac{-3m^2}{16\pi^2f^2}
\biggl(5\ln\frac{m^2}{\mu^2} +1\biggr) - \frac{3\Delta f}{f} +
\frac{\Delta m^2}{m^2} \right], \ee

\noindent for $K\to\pi\pi$,$m_K = m_\pi =m$, and

\bea \bra\pi^+\pi^- | {\cal O}^{(27,1),(3/2)} | K^0\ket_{log} & =
& -3i \frac{\alpha_{27}}{f^3} m^2_K \left[
\frac{-m^2_K}{12}\frac{1}{16\pi^2f^2} \biggl(114\ln
\frac{m^2_K}{\mu^2} +31
\ln5 -148 \ln2\right. \nonumber \\
& &  -16\cot^{-1} 2+46\biggr) - \frac{\Delta f_K}{f} -
\frac{2\Delta f_\pi}{f} \nonumber \\ & & \left.
+\frac{4}{3m^2_K}\biggl(\Delta m^2_K-\Delta m^2_\pi\biggr)
\right], \eea

\noindent for $K\to\pi\pi$, $m_{K(1-loop)} = 2m_{\pi(1-loop)}$.
The logarithmic corrections to the physical $\Delta I=3/2$
$K\to\pi\pi$ amplitude (included for completeness) are given by

\bea \bra \pi^+\pi^-|{\cal O}^{(27,1),(3/2)}|K^0\ket_{log}&=& -4i
\frac{\alpha_{27}}{f_Kf^2_\pi} \left[ -\frac{1}{12} m^4_K \left(
1-\frac{m^2_K}{m^2_\pi}\right) B(m^2_\pi, m^2_K, m^2_\eta)\right.
\nonumber \\
& & + m^2_K \left(\frac54 \frac{m^4_K}{m^2_\pi} -
\frac{13}{4}m^2_K +
2m^2_\pi\right) B(m^2_\pi, m^2_K, m^2_\pi) \nonumber \\
& & + \biggl(m^4_K-3m^2_\pi m^2_K +2m^4_\pi\biggr) B(m^2_K, m^2_\pi, m^2_\pi)
\nonumber \\
& & - \frac14 m^2_K \left(\frac{m^2_K}{m^2_\pi}+3\right)
A(m^2_\eta) + \left( \frac{-m^4_K}{m^2_\pi} - 4m^2_K
+4m^2_\pi\right) A(m^2_K)
\nonumber \\
& &\left. + \left( \frac54 \frac{m^4_K}{m^2_\pi} - \frac{45}{4}
m^2_K + 12 m^2_\pi\right) A(m^2_\pi)\right],\eea

\noindent where the imaginary part of expression (C5) is given by

\be Im(i\bra \pi^+\pi^-|{\cal O}^{(27,1),(3/2)} | K^0\ket)=
-\frac{2\alpha_{27}}{f_Kf^2_\pi}\frac{1}{16\pi f^2}
\sqrt{1-4\frac{m^2_\pi}{m^2_K}} (m^2_K-m^2_\pi) (m^2_K-2m^2_\pi
).\ee

\bigskip
\bigskip

\appendix
\section*{\bf APPENDIX D}
\setcounter{equation}{0} \setcounter{section}{1}
\renewcommand{\theequation}{D\arabic{equation}}
\bigskip

The logarithmic corrections for the quantities relevant for the
determination of the [(8,1)+(27,1)], $\Delta I=1/2$ $K\to\pi\pi$
amplitudes (as well as the pure (8,1) amplitudes, neglecting the
(27,1) expressions) are given by

\bea \bra 0|{\cal O}^{(8,1)}|K^0\ket_{log}\!\!\!\! & = &\!\!\!\!
\frac{4i\alpha_2}{f} (m^2_K-m^2_\pi) \left[ -\frac56 A(m^2_\eta)
-3A(m^2_K)
-\frac32A(m^2_\pi) -\frac{\Delta f_K}{f}\right] \nonumber \\
&& - \frac{4i\alpha_1}{f} \left[ \frac16\biggl(m^2_\pi-4m^2_K\biggr)
A(m^2_\eta) - m^2_KA(m^2_K) + \frac32 m^2_\pi A(m^2_\pi) \right],\nonumber \\
&& \eea

\be \bra0|{\cal O}^{(27,1)}|K^0\ket_{log} =
\frac{6i\alpha_{27}}{f} [ (m^2_\pi-4m^2_K) A(m^2_\eta) +4m^2_K
A(m^2_K) -m^2_\pi A(m^2_\pi)], \ee

\bea \bra \pi^+|{\cal O}^{(27,1),(1/2)} |K^+\ket_{\log} & = &
-\frac{4\alpha_{27}}{f^2} \left[ \frac18 \biggl(
-\frac{(7m^2_K-m^2_\pi)(m^2_K-m^2_\pi)^2}{q^2} + 7m^4_K
-6m^2_Km^2_\pi\right. \nonumber \\
& &  -m^4_\pi  -36(p_K\cdot p_\pi)^2 - 6(3m^2_K-5m^2_\pi)p_K\cdot
p_\pi \biggr)
B(q^2,m^2_K, m^2_\eta) \nonumber \\
& & +\frac18 \biggl(\frac{3(m^2_K+m^2_\pi)
(m^2_K-m^2_\pi)^2}{q^2} \nonumber\\
& & -3(m^2_K-m^2_\pi)^2 +20(p_K\cdot p_\pi)^2 - 6(m^2_K+m^2_\pi)
p_K\cdot p_\pi\biggr) B(q^2, m^2_K, m^2_\pi) \nonumber \\
& & +\frac38 \left(\frac{7m^4_K -8m^2_Km^2_\pi+m^4_\pi}{q^2} +
9m^2_K-m^2_\pi -10p_K\cdot p_\pi\right) A(m^2_\eta)\nonumber \\
& & - \left( \frac{3m^2_K (m^2_K-m^2_\pi)}{q^2} + 3m^2_K +10
p_K\cdot
p_\pi \right) A(m^2_K) \nonumber \\
& & +\frac38 \left(\frac{m^4_K-m^4_\pi}{q^2} - m^2_K + m^2_\pi -
6p_k\cdot p_\pi \right) A(m^2_\pi) \nonumber \\ &&\left. - \biggl(
\frac{\Delta f_K}{f} + \frac{\Delta f_\pi}{f}\biggr) p_K \cdot
p_\pi \right], \eea

\bea \bra\pi^+ | {\cal O}^{(8,1),(1/2)} |K^+ \ket_{log} & = &
\frac{4\alpha_1}{f^2} \left[\frac{1}{72} \biggl(
\frac{(7m^2_K-m^2_\pi)(m^2_K-m^2_\pi)^2}{q^2} - 7m^4_K
+6m^2_Km^2_\pi +
m^4_\pi \right.\nonumber \\
& & +36 (\pkppi)^2 + 6(3m^2_K-5m^2_\pi) \pkppi\biggr)
B(q^2,m^2_K,m^2_\eta)
\nonumber \\
& & +
\frac18 \biggl( \frac{3(m^2_K+m^2_\pi)(m^2_k-m^2_\pi)^2}{q^2}\nonumber \\
& & -3(m^2_K-m^2_\pi)^2 +20 (\pkppi)^2 - 6(m^2_K+m^2_\pi) \pkppi\biggr)
B(q^2,m^2_K, m^2_\pi) \nonumber \\
& & + \frac{1}{24} \left(\frac{-7m^4_K+8m^2_Km^2_\pi -
m^4_\pi}{q^2} -
9m^2_K+m^2_\pi - 30\pkppi\right) A(m^2_\eta) \nonumber \\
& & -\frac{1}{12} \left( \frac{m^4_K+4m^2_Km^2_\pi -5m^4_\pi}{q^2}
+11m^2_K + 5m^2_\pi + 30 \pkppi\right) A(m^2_K) \nonumber \\
& & \left. +\frac38 \left( \frac{m^4_K-m^4_\pi}{q^2} -
m^2_K+m^2_\pi -6\pkppi\right) A(m^2_\pi) - \biggl(\frac{\Delta
f_K}{f} + \frac{\Delta
f_\pi}{f} \biggr) \pkppi \right] \nonumber \\
& & -\frac{4\alpha_2}{f^2} m^2_K \left[ \frac{1}{12} \left(
\frac{(m^2_K-m^2_\pi)^2}{q^2} -m^2_K -m^2_\pi +6\pkppi\right)
B(q^2,
m^2_K, m^2_\eta) \right. \nonumber \\
& & +\frac14 \left( \frac{3(m^2_K-m^2_\pi)^2}{q^2} -
3(m^2_K+m^2_\pi)
+10\pkppi \right) B(q^2, m^2_K, m^2_\pi)\nonumber \\
& & -\frac{1}{12} \left(\frac{3(m^2_K-m^2_\pi)}{q^2} + 7\right)
A(m^2_\eta) -\frac12 \left(\frac{m^2_K - m^2_\pi}{q^2} +5\right)
A(m^2_K) \nonumber
\\
& & \left. + \frac34 \left(\frac{m^2_K-m^2_\pi}{q^2} - 3\right)
A(m^2_\pi) - \frac{\Delta f_k}{f} - \frac{\Delta f_\pi}{f}
\right], \eea

\be \bra\pi^+\pi^- | {\cal O}^{(27,1),(1/2)} | K^0\ket_{log} = -8i
\frac{\alpha_{27}}{f^3} m^2 \left[ -3m^2 \frac{1}{16\pi^2f^2}
\biggl( 5\ln \frac{m^2}{\mu^2} +1 \biggr) -\frac{3\Delta f}{f} +
\frac{\Delta m^2}{m^2} \right], \ee

\noindent for $K\to\pi\pi$, $m_K=m_\pi=m$,

\bea \bra\pi^+\pi^- | {\cal O}^{(8,1),(1/2)} | K^0\ket_{log} & = &
8i \frac{\alpha_1}{f^3} m^2 \left[ -\frac16 m^2
\frac{1}{16\pi^2f^2} \biggl(50\ln \frac{m^2}{\mu^2} -37\biggr) -
\frac{3\Delta f}{f} + \frac{\Delta m^2}{m^2} \right] \nonumber \\
& & +4i \frac{\alpha_2}{f^3} m^2 \left[ -2m^2 \frac{1}{16\pi^2f^2}
\biggl( \frac{101}{9} \ln \frac{m^2}{\mu^2} -\frac{47}{9}\biggr)
\right.
\\
& & \left. -\frac{3\Delta f}{f}+\frac{\Delta m^2}{m^2} +\frac{32
m^2}{f^2} (2L_1 + 2L_2 + L_3 +2L_4 +2L_6 +L_8)\right], \nonumber
\eea

\noindent for $K\to\pi\pi$, $m_K=m_\pi=m$,

\bea \bra\pi^+\pi^- | {\cal O}^{(27,1),(1/2)} | K^0\ket_{log} & =
& -3i \frac{\alpha_{27}}{f^3} m^2_K \left[ \frac{-m^2_K}{24}
\frac{1}{16\pi^2f^2} \biggl(456 \ln \frac{m^2_K}{\mu^2} - 20
(5+\ln 16) \right.
\nonumber \\
& &  -7\ln 5 + 232 \cot^{-1} 2\biggr) - \frac{\Delta f_K}{f} -
\frac{2\Delta f_\pi}{f} \nonumber \\ & & \left.
+\frac{4}{3m^2_K}\biggl(\Delta m^2_K-\Delta m^2_\pi\biggr)
\right], \eea

\noindent for $K\to\pi\pi$, $m_{K(1-loop)}=2m_{\pi(1-loop)}$, and

\bea \bra\pi^+\pi^- | {\cal O}^{(8,1),(1/2)} | K^0\ket_{log} & = &
3i \frac{\alpha_1}{f^3} m^2_K \left[ \frac{m^2_K}{72}
\frac{1}{16\pi^2f^2} \biggl(-518 \ln \frac{m^2_K}{\mu^2} -209 \ln
5 + 700 \ln 2
\right. \nonumber \\
& &  + 184 \cot^{-1} 2 +80 \biggr) - \frac{\Delta f_K}{f} -
\frac{2\Delta f_\pi}{f}\nonumber \\  & & \left.
+\frac{4}{3m^2_K}\biggl(\Delta m^2_K-\Delta m^2_\pi\biggr) \right]\nonumber \\
& & + 12i \frac{\alpha_2}{f^5} m^4_K (4L_4 - L_5 +8L_6 +4L_8),
\eea

\noindent for $K\to\pi\pi$, $m_{K(1-loop)}=2m_{\pi(1-loop)}$.  The
logarithmic corrections to the physical $\Delta I=1/2$
$K\to\pi\pi$ amplitude are given by

\bea \bra\pi^+\pi^- | {\cal O}^{(27,1),(1/2)} | K^0\ket_{log} & =
& -4i \frac{\alpha_{27}}{f_Kf^2_\pi} \left[ -\frac{2}{3} m^4_K
\left(
\frac{m^2_K}{m^2_\pi} -1\right) B(m^2_\pi, m^2_K, m^2_\eta)\right. \nonumber \\
& & +m^2_K \left(\frac{m^4_K}{2m^2_\pi} -\frac52 m^2_K +2m^2_\pi
\right) B(m^2_\pi, m^2_K, m^2_\pi) \nonumber \\
& & + \biggl(-2m^4_K +3m^2_Km^2_\pi - m^4_\pi\biggr) B(m^2_K, m^2_\pi, m^2_\pi)\nonumber \\
& & + m^2_\pi \biggl(m^2_K-m^2_\pi\biggr) B(m^2_K, m^2_\eta, m^2_\eta) +
\left( \frac{2m^4_K}{m^2_\pi} -\frac{15}{2} m^2_K + \frac92
m^2_\pi\right)
A(m^2_\eta)\nonumber \\
& & + \left( \frac{-5m^4_K}{2m^2_\pi} - \frac{11}{2} m^2_K +
10m^2_\pi\right ) A(m^2_K)\nonumber \\ & &\left. +
\left(\frac{m^4_K}{2m^2_\pi} - 3m^2_K
 +\frac32 m^2_\pi\right) A(m^2_\pi)\right],
\eea

\bea \bra\pi^+\pi^- | {\cal O}^{(8,1),(1/2)} | K^0\ket_{log} & = &
4i \frac{\alpha_{1}}{f_Kf^2_\pi} \left[\frac16 m^4_K
\left(\frac{m^2_K}{m^2_\pi}-1\right) B(m^2_\pi, m^2_K,
m^2_\eta)\right.
\nonumber \\
& & + \frac12 m^2_K \left(\frac{m^4_K}{m^2_\pi} -5m^2_K
+4m^2_\pi\right) B(m^2_\pi, m^2_K, m^2_\pi) \nonumber \\
& & - \biggl(2m^4_K - 3m^2_Km^2_\pi + m^4_\pi\biggr) B(m^2_K, m^2_\pi, m^2_\pi) \\
& & -\frac19 m^2_\pi \biggl(m^2_K-m^2_\pi\biggr) B(m^2_K, m^2_\eta, m^2_\eta)
- \frac12
\left( \frac{m^4_K}{m^2_\pi} +m^2_\pi\right) A(m^2_\eta) \nonumber \\
& & + \biggl(5m^2_\pi - 3m^2_K\biggr) A(m^2_K) \nonumber \\
& &\left. + \frac12 \left(\frac{m^4_K}{m^2_\pi} - 6m^2_K
+3m^2_\pi\right) A(m^2_\pi) \right]\nonumber \\ & & +
\frac{64i}{f^5} \alpha_2 (m^2_K-m^2_\pi)[2m^2_K L_4  + (-4 L_4-L_5
+8L_6 +4L_8) m^2_\pi ]. \nonumber \eea

\noindent The imaginary parts of expressions (D9) and (D10) are
given by

\be Im(i\bra\pi^+\pi^- | {\cal O}^{(27,1),(1/2)} | K^0\ket)=
\frac{2\alpha_{27}}{f_Kf^2_\pi}\frac{1}{16\pi f^2}
\sqrt{1-4\frac{m^2_\pi}{m^2_K}} (m^2_K-m^2_\pi) (2m^2_K
-m^2_\pi),\ee

\be Im(i\bra\pi^+\pi^- | {\cal O}^{(8,1),(1/2)} | K^0\ket)=
-\frac{2\alpha_{1}}{f_Kf^2_\pi}\frac{1}{16\pi f^2}
\sqrt{1-4\frac{m^2_\pi}{m^2_K}} (m^2_K-m^2_\pi)
(2m^2_K-m^2_\pi).\ee


\begin{thebibliography}{99}

\bibitem{noaki}  CP-PACS Collaboration, J.I. Noaki \etal,
hep-lat/0108013.   

\bibitem{blum}  RBC Collaboration, T. Blum \etal, hep-lat/0110075.   

\bibitem{fanti}  NA48 Collaboration, V. Fanti \etal, Phys.\ Lett.\ B {\bf 465}, 335
(1999).   

\bibitem{alavi}  KTeV Collaboration, A. Alavi-Harati \etal, Phys.\ Rev.\ Lett.\ {\bf83},
22 (1999).  

\bibitem{pk} For an attempt at $K \rightarrow \pi \pi$ with staggered fermions and
using lowest order ChPT see, D.Pekurovsky and G.Kilcup, Phys.\
Rev.\ D {\bf 64}, 074502 (2001).

\bibitem{gm1} For a recent theoretical review see, G. Martinelli,
Nucl.\ Phys.\ B (Proc. Suppl.) {\bf 106}, 98 (2002).

\bibitem{bern} C. Bernard, T. Draper, A. Soni, H.D. Politzer, and
M.B. Wise, Phys.\ Rev.\ D {\bf 32}, 2343 (1985).   

\bibitem{maiani} L. Maiani and M. Testa, Phys.\ Lett.\ B {\bf 245}, 585
(1990).  

\bibitem{cirig}  V. Cirigliano and E. Golowich, Phys.\ Lett.\ B {\bf
475}, 351 (2000); Phys.\ Rev.\ {\bf D65}, 054014 (2002).   

\bibitem{bijnens} J. Bijnens, E. Pallante, and J. Prades, Nucl.\
Phys.\ B {\bf 521}, 305 (1998).   

\bibitem{berntwo} C. Bernard, T. Draper, G. Hockney and A. Soni, Nucl.\
Phys.\ B (Proc. Suppl.) {\bf4}, 483 (1988).   


\bibitem{dawson} C. Dawson {\it et al}., Nucl.\ Phys.\ B {\bf 514}, 313
(1998).   

\bibitem{spqcdr} For an ongoing attempt with Wilson fermions
at $O(p^4)$ see \cite{gm1} and also for a specific recipe for the
$\Delta I=3/2$ case, see Ph.Boucaud et. al. [SPQcdR Collab],
Nucl.\ Phys.\ B (Proc. Suppl.) {\bf106}, 329 (2002).


\bibitem{ciuch} M. Ciuchini, E. Franco, G. Martinelli, L. Reina and L.
Silvestrini, Z. Phys.\ C {\bf 68}, 239 (1995).   

\bibitem{bucha} G. Buchalla, A.J. Buras, and M.E. Lautenbacher, Rev.\
Mod.\ Phys.\ {\bf68}, 1125 (1996).   

\bibitem{georgi} For a  succinct introduction, see, H. Georgi, \textit{Weak
Interactions and Modern Particle Theory}
(Benjamin/Cummings, New York, 1984).  

\bibitem{golt} See, e.g., M. Golterman and E. Pallante hep-lat/0110206.

\bibitem{gass} J. Gasser and H. Leutwyler, Ann.\ Phys.\ (N.Y.) {\bf158}, 142
(1984); Nucl.\ Phys.\ B {\bf 250}, 465 (1985).   

\bibitem{kambor} J. Kambor, J. Missimer and D. Wyler, Nucl.\ Phys.\
{\bf B346}, 17 (1990).  

\bibitem{ecker} G. Ecker, J. Kambor and D. Wyler, Nucl.\ Phys.\ {\bf
B394}, 101 (1993). See also, E. Pallante, A. Pich and I. Scimemi,
Nucl.\ Phys.\ {\bf B617}, 441 (2001).   

\bibitem{dono} J. Donoghue, E. Golowich and B. Holstein, Phys.\ Lett.\
{\bf B119}, 412 (1982).   

\bibitem{spqcdr2} Note also that in \cite{spqcdr} a different recipe
for the $\Delta I=3/2$ is used than the one advocated here;
they are inserting momentum in $K \rightarrow \pi \pi$
whereas in our case momentum insertion is required
only for $K \rightarrow \pi$. In principle, using momentum
insertion for both cases could
enhance redundacy even further although momentum insertion
in $K \rightarrow \pi \pi$ may well be somewhat harder.

\bibitem{golttwo} M. Golterman and E. Pallante, hep-lat/0108029.  

\bibitem{lell} L. Lellouch and M. Luscher, Commun.\ Math.\ Phys.\ {\bf 219}, 31 (2001)
 ; see also C.J.D.Lin, G. Martinelli, C. T. Sachrajda and M. Testa,
Nucl.\ Phys.\ B (Proc. Suppl.) {\bf 109}, 218 (2002).

\bibitem{buch} M. Buchler, G. Colangelo, J. Kambor and F. Orellana,
Phys.\ Lett.\ B {\bf 521}, 22 (2001); G. Colangelo, Nucl.\ Phys.\
B (Proc. Suppl.) {\bf 100}, 53 (2002).   

\bibitem{kamb}  Joachim Kambor, Ph.D thesis, Swiss Federal
Institute of Technology, 1990.   

\bibitem{wolfe} Carl Wolfe, Ph.D Thesis, York University,
1999   

\bibitem{mert} R. Mertig, M. Bohm, and A. Denner, Comput.
Phys. Commun. {\bf 64}, 345, (1991).

\bibitem{wolfram} S. Wolfram, \textsc{Mathematica} - \emph{a system for Doing
Mathematics by Computer} (Addison-Wesley, New York, 1988).

\bibitem{pass} G. Passarino and M. Veltman, Nucl.\ Phys.\ {\bf
B160}, 151 (1979).


\end{thebibliography}
\end{document}